\pdfoutput=1

\documentclass[11pt]{article}

\usepackage[preprint]{acl}

\usepackage{times}
\usepackage{latexsym}

\usepackage[T1]{fontenc}

\usepackage[utf8]{inputenc}

\usepackage{microtype}

\usepackage{inconsolata}

\usepackage{graphicx}
\usepackage{graphicx}
\usepackage{amssymb}
\usepackage{bm}
\usepackage{amsmath}
\usepackage{algorithm}
\usepackage{algorithmic}
\usepackage{booktabs}
\usepackage{multirow}
\usepackage{multicol}
\usepackage{pifont}
\usepackage{enumitem}
\usepackage{colortbl}
\usepackage{tabularx}
\usepackage{makecell}
\definecolor{Black}{rgb}{0,0,0}
\newcommand{\blue}[1]{$_{\color{Black}\downarrow #1}$}
\newcommand{\red}[1]{$_{\color{Black}\uparrow #1}$}
\newcommand{\obs}[1]{\textit{\textbf{#1}}}
\definecolor{tableheadcolor}{RGB}{255,255,255}

\title{Multifaceted Evaluation of Audio-Visual Capability for MLLMs: Effectiveness, Efficiency, Generalizability and Robustness}

\author{
Yusheng Zhao\textsuperscript{$\heartsuit$}, Junyu Luo\textbf{\textsuperscript{$\heartsuit$}}, Xiao Luo\textsuperscript{\ding{171}}, Weizhi Zhang\textsuperscript{$\diamondsuit$},   \\ \textbf{Zhiping Xiao}\textsuperscript{\ding{168}}, \textbf{Wei Ju}\textsuperscript{$\heartsuit$}, \textbf{Philip S. Yu}\textsuperscript{$\diamondsuit$}, \textbf{Ming Zhang}\textsuperscript{$\heartsuit$} \\
\textsuperscript{$\heartsuit$} Peking University \quad \textsuperscript{\ding{171}} University of California, Los Angeles \\
\textsuperscript{$\diamondsuit$} University of Illinois Chicago \quad \textsuperscript{\ding{168}} University of Washington \\
\texttt{\{yusheng.zhao, luojunyu\}@stu.pku.edu.cn}, 
\texttt{xiaoluo@cs.ucla.edu},\\
\texttt{\{wzhan42, psyu\}@uic.edu},
\texttt{patxiao@uw.edu}, 
\texttt{\{juwei,mzhang\_cs\}@pku.edu.cn}
}

\begin{document}
\maketitle
\begin{abstract}
Multi-modal large language models (MLLMs) have recently achieved great success in processing and understanding information from diverse modalities (\emph{e.g.}, text, audio, and visual signals). Despite their growing popularity, there remains a lack of comprehensive evaluation measuring the audio-visual capabilities of these models, especially in diverse scenarios (\emph{e.g.}, distribution shifts and adversarial attacks). 
In this paper, we present a multifaceted evaluation of the audio-visual capability of MLLMs, focusing on four key dimensions: \textit{\textbf{effectiveness}}, \textit{\textbf{efficiency}}, \textit{\textbf{generalizability}}, and \textit{\textbf{robustness}}.
Through extensive experiments, we find that MLLMs exhibit strong zero-shot and few-shot generalization abilities, enabling them to achieve great performance with limited data.
However, their success relies heavily on the vision modality, which impairs performance when visual input is corrupted or missing.
Additionally, while MLLMs are susceptible to adversarial samples, they demonstrate greater robustness compared to traditional models.
The experimental results and our findings provide insights into the audio-visual capabilities of MLLMs, highlighting areas for improvement and offering guidance for future research.
\end{abstract}

\section{Introduction}
\label{sec:intro}
Multi-modal large language models (MLLMs) \cite{lin2023video, zhang2023video, cheng2024videollama, fu2024vita, wu2024next, jin2024chat, zhang2024mm} have shown impressive performance in processing and understanding information from multiple modalities, such as text, image, and audio. 
The prevalent paradigm of MLLMs involves using modality-specific encoders \cite{tan2019lxmert, ando2023use} to process individual modalities (e.g., image, video, and audio) into tokens, which are then fed into a large language model (LLM). Attention is computed across modalities, fusing information \cite{cheng2024videollama,fu2024vita}.
The success of these models enables a wide range of applications, including image captioning \cite{bucciarelli2024personalizing, zhang2024differential}, visual question answering \cite{kuang2024natural, xu2024mlevlm, zhao2025lova3}, and multi-modal scene understanding \cite{luo2024delving, fan2024mllm, xiong20253ur}.

Among the modalities in the real world, text, vision, and audio are particularly important due to their prevalence and richness of information \cite{qi2000integrating, li2018read}. Therefore, evaluating the audio-visual capability of MLLMs is crucial for understanding their overall performance and potential applications in real-world scenarios \cite{geng2023dense, chen2024real}.
However, previous evaluation efforts \cite{bai2023touchstone, xu2024lvlm, chen2024we, kahng2024llm} have mostly focused on vision and language modalities, often ignoring the audio modality. This oversight limits our understanding of the full potential and limitations of MLLMs, especially in scenarios where audio information plays a critical role \cite{lyu2023macaw, ye2024cat}.
For example, in autonomous driving, audio signals such as sirens and horns are crucial for safety \cite{sun2021emergency, furletov2021auditory}. In multimedia content analysis, audio cues are essential for understanding context and emotions \cite{liu2024harnessing, qi2024movie}.

Compared to previous efforts involving only visual and linguistic modalities \cite{hu2024bliva, pi2024perceptiongpt, li2024dtllm}, the inclusion of the audio modalities poses several challenges. 
Firstly, there are differences in the informativeness of different modalities \cite{evangelopoulos2009video, wang2014adults, fan2023pmr}. Visual clues are often more informative (\emph{e.g.}, recognizing human actions or understanding locations), while audio signals can be more informative in rarer situations (\emph{e.g.}, detecting fire alarms or musical instruments). The multi-modal learning system may rely on the dominant modality (\emph{i.e.}, vision) while disregarding information from the other (\emph{i.e.}, audio) \cite{fan2024overcome, wu2025balanced}. 
Secondly, the audio and visual modalities are complementary \cite{ma2022multimodal, gungor2023complementary}. When one modality is corrupted or missing, the other can provide supplementary information to aid scene understanding. The audio-visual LLMs should be able to leverage the complementary information from both modalities effectively. 
Thirdly, the audio modality is noisier and less structured than the visual modality \cite{gao2021visualvoice,  liu2022audio}, as audio signals are often affected by background noise \cite{moncrieff2007online}, reverberation \cite{usher2007enhancement}, and other distortions \cite{preis1982phase}. Although there are some related works of audio-visual evaluation \cite{tseng2024av, wang2024audiobench, sung2024avhbench}, they have mostly focused on effectiveness, whereas this work is more comprehensive, focusing on various aspects of MLLMs' ability.

In this paper, we focus on evaluating the audio-visual capability of MLLMs. Specifically, we aim to provide a comprehensive evaluation of their audio-visual capability across four key dimensions: \textit{\textbf{\ding{182} Effectiveness}},  measured by performance using audio and/or visual inputs. \textit{\textbf{\ding{183} Efficiency}}, which includes both data efficiency (how the models perform under limited data) and computational efficiency (\emph{e.g.}, model size, memory consumption and inference speed). \textit{\textbf{\ding{184} Generalizability}}, focusing on performance under test-time distribution shifts. \textit{\textbf{\ding{185} Robustness}}, which measures resilience against adversarial perturbations.

We conduct extensive experiments around the four aforementioned aspects with several observations. Firstly, MLLMs are generally competitive in understanding audio-visual information, although they rely heavily on the visual modality. Secondly, their over-reliance on the visual modality leads to poor performance when the video inputs are under test-time distribution shifts. Thirdly, the MLLMs exhibit high data efficiency, achieving superior performance under limited data. However, they lag behind traditional models in terms of computational efficiency. Fourthly, MLLMs can be fooled by adversarial samples, but they are more robust compared to traditional models.

The contribution of this work is summarized as follows: (1) We establish a thorough evaluation framework of the audio-visual capability of MLLMs by considering four crucial dimensions: effectiveness, efficiency, generalizability, and robustness. (2) Extensive experiments reveal that MLLMs exhibit strong zero-shot and few-shot audio-visual capabilities, despite their over-reliance on the visual modality, which hinders their performance under test-time distribution shifts in vision. (3) The experiments also reveals that MLLMs are more robust against adversarial perturbations compared to traditional models.

\begin{figure*}
    \centering
    \includegraphics[width=\linewidth]{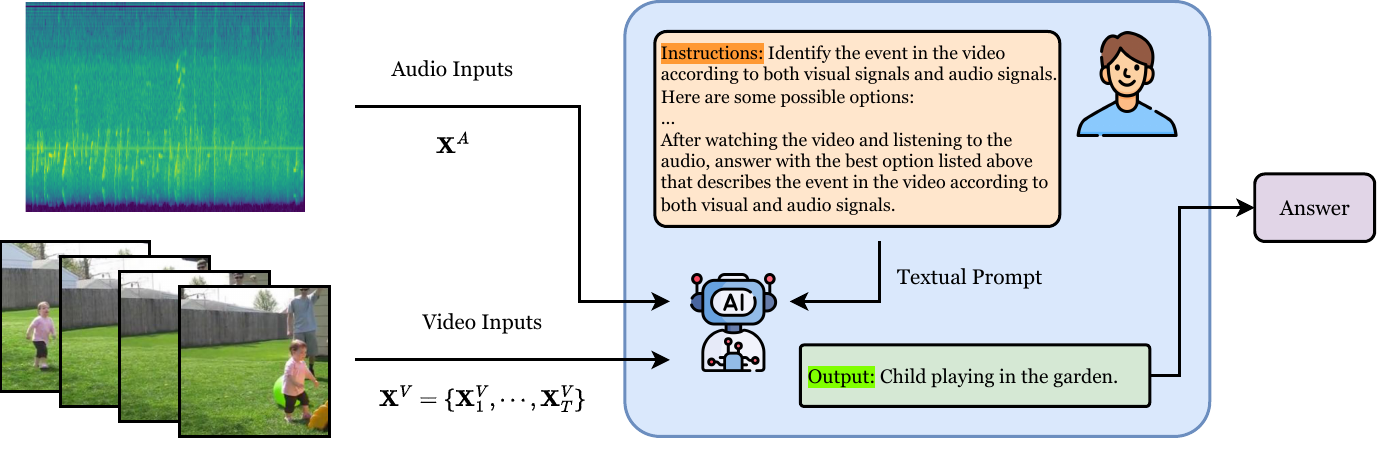}
    \vspace{-5mm}
    \caption{The framework of our evaluation of audio-visual capabilities of MLLMs. The MLLM takes audio signals, video frames and textual instructions as inputs and generates the corresponding output.}
    \vspace{-2mm}
    \label{fig:framework}
\end{figure*}
\section{Related Works}
\subsection{Multi-modal Large Language Models}
Multi-modal large language models (MLLMs) \cite{hu2024bliva, fei2024multimodal, zhan2024anygpt,fu2025vita} integrate information from multiple modalities, such as text, images, and audio, to improve understanding and generation capabilities. These models leverage the strengths of each modality by encoding the knowledge with modality-specific encoders \cite{gong2021ast, arnab2021vivit, han2022survey} and fusing the multi-modal tokens with large language models \cite{touvron2023llama, yang2024qwen2}. Recent advancements in MLLMs have shown significant improvements in their visual and linguistic abilities, allowing large language models to recognize visual inputs such as images and videos \cite{lin2024vila, pi2024perceptiongpt}. Nevertheless, in real-world scenarios, audio signals are sometimes crucial for understanding the context of the input, with several works focusing on audio-visual large language models \cite{zhang2023video,cheng2024videollama, fu2025vita}. In this work, we provide a comprehensive evaluation of these models, measuring their effectiveness, efficiency, generalizability and robustness.

\subsection{Test-time Distribution Shift}
Test-time distribution shift is a common challenge in real-world applications, where the test data distribution differs from the training distribution, leading to a significant drop in model performance \cite{darestani2022test, sinha2023test, liang2025comprehensive}. To mitigate the problem during test time, test-time adaptation methods have been proposed to adapt the model during test time without accessing the training data \cite{boudiaf2022parameter, chen2022contrastive, yuan2023robust}. However, these methods are often computationally expensive and assume simple classification tasks \cite{niu2022efficient, lee2023towards, lee2024entropy}, limiting their applicability to multi-modal large language models. In this work, we investigate the generalizability of multi-modal large language models to test-time distribution shift.

\subsection{Adversarial Robustness}
Adversarial robustness is a critical aspect of deep neural networks, ensuring that models are robust to adversarial samples \cite{szegedy2013intriguing, moosavi2016deepfool, chakraborty2018adversarial}. Adversarial samples are specially designed inputs to fool the model into making wrong predictions. The robustness of multi-modal large language models against adversarial samples is important for safety-related real-world applications, including autonomous driving \cite{cui2024survey}, robotics \cite{elmallah2024human}, and finance \cite{gan2024mme, xue2024famma}. In this work, we evaluate the robustness of audio-visual MLLMs against adversarial attacks, providing insight about the reliability of these models.

\section{The Evluation}
\subsection{Problem Definition}
In the evaluation of the audio-visual capabilities of MLLMs, we denote the visual input (\emph{i.e.} the video) as $\bm X^V$, consisting a sequence of frames $\{\bm X^V_1, \bm X^V_2, \cdots, \bm X^V_T\}$, and the audio input as $\bm X^A$. Given the textual instruction of $I$, the MLLM model $\mathcal M$ is expected to generate the output string denoted as $O=\mathcal M(\bm X^V, \bm X^A, I)$. The generated output is then compared with the ground truth output $O^*$ to evaluate the performance of the model.

\subsection{Compared Methods}
We adopt two popular MLLMs, \emph{i.e.} VideoLLaMA 2 \cite{cheng2024videollama} and VITA 1.5 \cite{fu2025vita}. VideoLLaMA 2 is a state-of-the-art MLLM for video understanding, with video, audio and text as its inputs. VITA 1.5 is another multi-modal LLM designed for video understanding, which has good audio-visual capabilities. For these MLLMs, we also train a fine-tuned version on the dataset for the evaluation. For comparison with traditional audio-visual approaches, we also include a SOTA audio-visual classification model, CAV-MAE \cite{gong2023contrastive}, which is fine-tuned on the adopted datasets. When measuring the performance under test-time distribution shifts, we also include several test-time adaptation methods, including Tent \cite{wang2020tent}, MMT \cite{shin2022mm}, EATA \cite{niu2022efficient}, SAR \cite{niu2023towards}, READ \cite{yang2024test}, and ABPEM \cite{zhao2025attention}.

\begin{table*}[ht]
\centering
\resizebox{0.95\textwidth}{!}{
\begin{tabular}{l ccc c ccc}
\Xhline{1.2pt}
\rowcolor{tableheadcolor!20} & \multicolumn{3}{c}{Kinetics50} && \multicolumn{3}{c}{VGGSound} \\
\cline{2-4} \cline{6-8} \rowcolor{tableheadcolor!20} \multirow{-2}{*}{Models} & Overall & Video-Only & Audio-Only && Overall & Video-Only & Audio-Only \\
\Xhline{1.2pt}
\rowcolor{gray!10}CAV-MAE & \underline{82.3} & 67.0 & \textbf{46.0} && \textbf{65.5} & 26.4 & \textbf{51.9}  \\
VideoLLaMA (Zero-Shot) & 73.2\blue{9.1} & 76.5\red{9.5} & 14.3\blue{31.7} && 59.3\blue{6.2} & \textbf{49.1}\red{22.7} & 35.3\blue{16.6}   \\
\rowcolor{gray!10}VideoLLaMA (SFT) & 78.9\blue{3.4} & 76.6\red{9.6} & \underline{17.1}\blue{28.9} && \underline{63.1}\blue{2.4} & \textbf{49.1}\red{22.7} & \underline{44.1}\blue{7.8} \\
VITA (Zero-Shot) & 70.5\blue{11.8} & \underline{77.5}\red{10.5} & 7.6\blue{38.4} && 29.8\blue{35.7} & 32.6\red{6.2} & 2.5\blue{49.4}  \\
\rowcolor{gray!10}VITA (SFT) & \textbf{83.6}\red{1.3} & \textbf{84.3}\red{17.3} & 9.9\blue{36.1} && 32.0\blue{33.5} & \underline{43.0}\red{16.6} & 13.0\blue{38.9}  \\
\Xhline{1.2pt}
\end{tabular}}
\vspace{-1mm}
\caption{Overall effectiveness of visual-audio models. We \textbf{bold} the best results and \underline{underline} the second-best.}
\vspace{-4mm}
\label{tab:effective}
\end{table*}

\subsection{Datasets}
We adopt two basic datasets, \emph{i.e.} Kinetics50 \cite{kay2017kinetics, yang2024test} and VGGSound \cite{chen2020vggsound}. Based on these datasets, we adopt corrupted versions under test-time distribution shifts (\emph{i.e.} Kinetics50-C and VGGSound-C) to evaluate the generalizability of MLLMs. Moreover, we also construct the adversarial versions of these datasets, \emph{i.e.} Kinetics50-A and VGGSound-A, to evaluate the robustness of MLLMs against adversarial perturbations. The datasets used in this paper are described as follows.

\smallskip
\noindent\textbf{Kinetics50} \cite{kay2017kinetics, yang2024test} is a subset of the Kinetics dataset \cite{kay2017kinetics}, which contains 400 classes of human actions. The subset contains 50 randomly selected classes \cite{yang2024test}, composing of 29k training samples and 2.5k test samples. In this dataset, visual clues play a more important role than audio signals.

\smallskip
\noindent\textbf{VGGSound} \cite{chen2020vggsound} is a dataset for audio-visual classification, which contains 309 classes of the events. The dataset consists of 160k training video clips and 14k test video clips from YouTube. For this dataset, audio signals are relatively more informative than the visual modality. 

\smallskip
\noindent\textbf{Kinetics50-C and VGGSound-C} \cite{yang2024test} are corrupted versions of Kinetics50 and VGGSound, respectively. The corrupted versions are constructed by adding different types of corruptions to the audio or visual inputs in the test set, making the test distributions different from the training ones. We adopt 15 types of corruptions for the visual modality and 6 types of corruptions for the audio modality following \citet{hendrycks2019benchmarking} and \citet{yang2024test}. 

\smallskip
\noindent\textbf{Kinetics50-A and VGGSound-A} are adversarial versions of Kinetics50 and VGGSound, respectively. They are constructed by adding adversarial perturbations to the visual inputs in the test set, making them adversarial samples. We adopt two commonly used adversarial attack methods, \emph{i.e.} Fast Gradient Sign Method (FGSM, proposed by \citet{goodfellow2014explaining}) and Projected Gradient Descent (PGD, proposed by \citet{madry2017towards}) to introduce the adversarial perturbations.

\section{Experiments and Analysis}
\subsection{Experimental Settings}
We adopt two state-of-the-art MLLM models, \emph{i.e.} VideoLLaMA \cite{cheng2024videollama} and VITA \cite{fu2025vita}. For VideoLLaMA, we use version 2.1, with Qwen 2 (7B) \cite{yang2024qwen2} as its language processor. For VITA, we use version 1.5. We also use supervised fine-tuning to obtain the fine-tuned versions of these models. All experiments are performed on NVIDIA A100 GPUs. In the evaluation, the results are reported in terms of percentage accuracy, unless otherwise specified.

\subsection{Effectiveness}
\label{sec:effective}
\textbf{Overall Effectiveness}
We first show the overall effectiveness of MLLMs in terms of their audio-visual capability. We evaluate the models' performance on Kinetics50 and VGGSound datasets, and the results are shown in Table \ref{tab:effective}. \obs{Observation 1: MLLMs demonstrate competitive audio-visual capability.} For Kinetics50, the MLLMs show performance comparable to the SOTA traditional model (\emph{i.e.} CAV-MAE), with the SFT version outperforming the zero-shot version. For VGGSound, VideoLLaMA still achieves comparable results with CAV-MAE, while VITA fails to reach the same level of performance. This discrepancy is due to the fact that, for the VGGSound dataset, the audio modality is more informative than the visual modality, and VITA relies heavily on the visual modality. Another reason (which we will elaborate on later in Section \ref{sec:case}, Case 2) is the confusion between speech and textual instructions.
\obs{Observation 2: MLLMs rely heavily on the visual modality}, which is demonstrated by the results when the visual signals are removed, as shown in Table \ref{tab:effective} (Audio-Only). During training, the two modalities are imbalanced, with vision being the dominant modality, a phenomenon observed in previous literature \cite{zhang2024modality, wu2025balanced}. This causes the model to rely heavily on vision during inference, while the audio is not fully utilized. This over-reliance on vision can be problematic when the audio modality carries important information, as we will show in Section \ref{sec:case}, Case 3.

\begin{figure*}
    \centering
    \includegraphics[width=\linewidth]{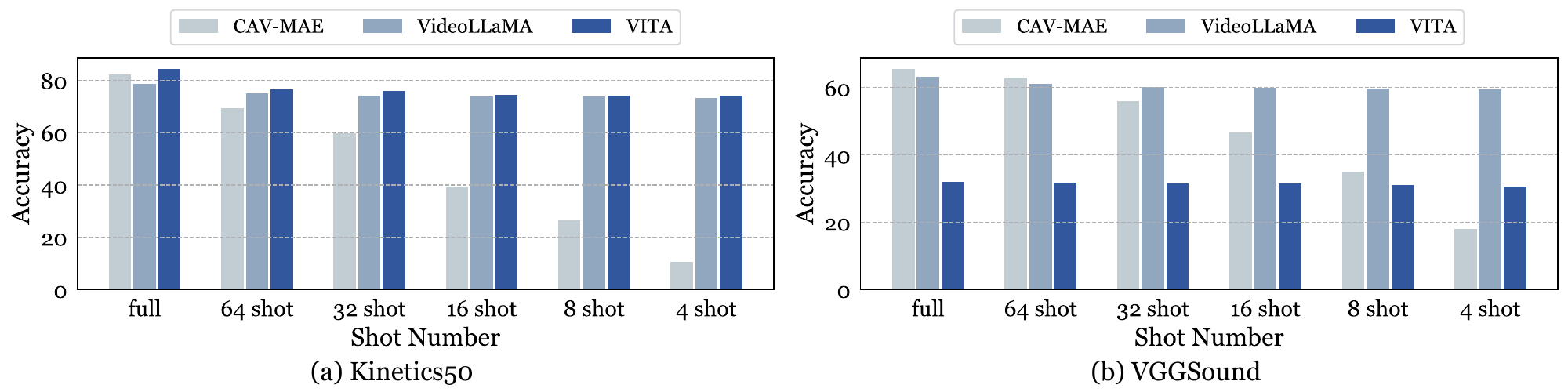}
    \vspace{-5mm}
    \caption{Data efficiency comparison of various models. We compare the models' performance under limited fine-tuning data, and show the results on the Kinetics50 (a) and VGGSound (b) datasets.}
    \vspace{-3mm}
    \label{fig:fewshot}
\end{figure*}

\smallskip
\noindent\textbf{Synergy of Visual and Audio Modalities}
We then provide an analysis of the synergy of visual and audio modalities. We evaluate the models with only one modality, and the results are shown in Table \ref{tab:effective} (Video/Audio-Only columns). \obs{Observation 3: when the MLLM cannot obtain enough information from one modality, there is little or no synergy between the modalities}, and the model's performance suffers as a result. In this case, when the MLLM cannot obtain enough information from the audio inputs, there is no synergy between the audio and video. This explains why, in some cases, the MLLM performs better when the audio input is removed (\emph{e.g.}, VITA on both datasets). On the other hand, when the MLLM can obtain sufficient information from both modalities, the synergy between the modalities can be observed, and the model's performance improves as a result (\emph{e.g.} VideoLLaMA on the VGGSound dataset).

\begin{table}[t]
    \centering
    \resizebox{0.47\textwidth}{!}{
    \begin{tabular}{l c c cc}
        \Xhline{1.2pt}
    \rowcolor{tableheadcolor!20}  &  &  & \multicolumn{2}{c}{Inference}\\
    \cline{4-5}\rowcolor{tableheadcolor!20}\multirow{-2}{*}{Models}  & \multirow{-2}{*}{Size} & \multirow{-2}{*}{Training Time} & Time & GPUMem\\
    \Xhline{1.2pt}
    \rowcolor{gray!10}CAV-MAE & 0.16B & 1.8h & 0.045s & 2.5GB  \\
    VideoLLaMA & 7B & 17h & 0.53s &  19GB  \\
    \rowcolor{gray!10}VITA & 7B & 16h & 0.58s & 19GB \\
    \Xhline{1.2pt}
    \end{tabular}}
    \vspace{-1mm}
    \caption{Models' computation efficiency comparison. Training time is measured in terms of GPU hours. Inference time is measured in terms of the time of processing one input sample. GPUMem is the GPU memory usage during inference. All experiments are conducted on the Kinetics50 dataset with NVIDIA A100 GPUs.}
    \vspace{-2mm}
    \label{tab:efficiency}
    \end{table}

\subsection{Efficiency}

\noindent\textbf{Computational Efficiency}
Next, we show the differences in computational resources of MLLMs compared to the traditional model, and the results are shown in Table \ref{tab:efficiency}. Specifically, we report the model size (measured by the number of parameters), the training time, the inference time, and the GPU memory usage during inference. The training and inference experiments are performed on the Kinetics50 dataset. During inference, we set the batch size to 1 for a fair comparison. \obs{Observation 4: MLLMs are less efficient in terms of computation.} As shown by the results, although MLLMs have larger model sizes, longer training times, and more inference computation compared to the traditional model, they can still achieve real-time inference on a single GPU, making them applicable in real-world scenarios. 

\smallskip
\noindent\textbf{Data Efficiency}
We then measure the models' data efficiency by evaluating their performance under limited fine-tuning data. We show the results on the Kinetics50 and VGGSound datasets in Figure \ref{fig:fewshot}, where we use few-shot training data to fine-tune the models and measure their accuracy. \obs{Observation 5: MLLMs have high data efficiency.} As shown by the results, MLLMs are generally data-efficient, and their performance drops only marginally when the amount of fine-tuning data is reduced (as demonstrated by a mild decrease from the full dataset to few-shot cases). In contrast, the traditional model (even with pretraining) suffers more from the lack of data. This demonstrates the superior audio-visual capability of MLLMs when the data is scarce.

\begin{figure}[t]
    \centering
    \includegraphics[width=\linewidth]{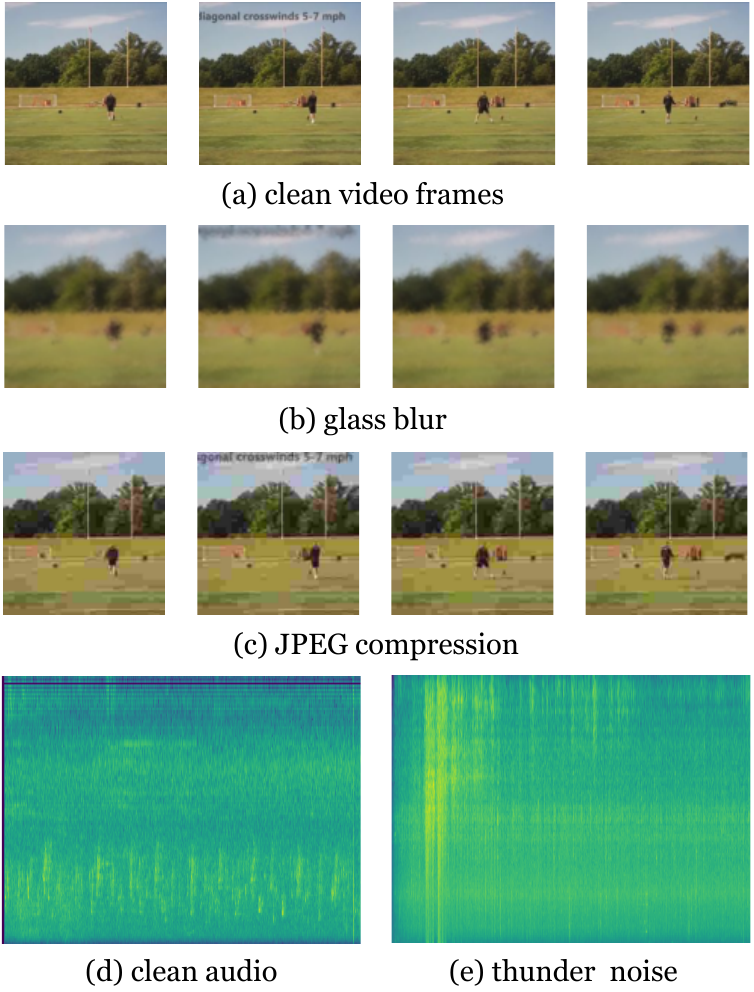}
    \vspace{-4mm}
    \caption{Visualization of input video frames and audio signals. The clean video frames and audio signals are shown in subfigures (a) and (d), while the corrupted versions are shown in subfigures (b), (c), and (e).}
    \vspace{-4mm}
    \label{fig:corruption-vis}
\end{figure}

\begin{table*}[ht]
\centering
\resizebox{\textwidth}{!}{
\begin{tabular}{l ccc c ccc c ccc c ccc c}
\Xhline{1.2pt}
\rowcolor{tableheadcolor!20} & \multicolumn{3}{c}{Noise} && \multicolumn{3}{c}{Weather} &  & \multicolumn{3}{c}{Noise} && \multicolumn{3}{c}{Weather} &  \\
\cline{2-4} \cline{6-8} \cline{10-12} \cline{14-16}\rowcolor{tableheadcolor!20}\multirow{-2}{*}{Models}
& Gauss. & Traff. & Crowd. && Rain & Thund. & Wind & \multirow{-2}{*}{Avg.} & Gauss. & Traff. & Crowd. && Rain & Thund. & Wind & \multirow{-2}{*}{Avg.}\\
\Xhline{1.2pt}
\rowcolor{gray!10}CAV-MAE & 73.7 & 65.5 & 67.9 && 70.3 & 67.9 & 70.3 & 69.3 & 37.0 & 25.5 & 16.8 && 21.6 & 27.3 & 25.5 & 25.6 \\
\quad +MMT & 70.8 & 69.2 & 68.5 && 69.0 & 69.8 & 68.5 & 69.4 & 14.1 & 5.2 & 6.4 && 9.8 & 8.6 & 4.5 & 7.6 \\
\rowcolor{gray!10}\quad +Tent & 73.9 & 67.4 & 68.5 && 70.4 & 66.5 & 70.4 & 69.6 & 10.6 & 2.6 & 1.8 && 2.3 & 3.3 & 4.1 & 4.5 \\
\quad +EATA & 73.7 & 66.1 & 68.5 && 69.5 & 70.6 & 69.4 & 69.4 & 39.2 & 26.1 & 22.9 && 26.0 & 31.7 & 30.4 & 29.4 \\ 
\rowcolor{gray!10}\quad +SAR & 73.7 & 65.4 & 68.2 && 69.9 & 67.2 & 70.2 & 69.1 & 37.4 & 9.5 & 11.0 && 12.1 & 26.8 & 23.7 & 20.1 \\ 
\quad +READ & \underline{74.1} & \underline{69.0} & \underline{69.7} && \underline{71.1} & \underline{71.8} & \underline{70.7} & \underline{71.1} & \underline{40.4} & \underline{28.9} & \underline{26.6} && \underline{30.9} & \underline{36.7} & \underline{30.6} & \underline{32.4} \\
\rowcolor{gray!10}\quad +{ABPEM} & \textbf{74.8} & \textbf{71.3} & \textbf{71.5} && \textbf{71.9} & \textbf{73.8} & \textbf{71.6} & \textbf{72.5} & \textbf{40.6} & \textbf{33.7} & \textbf{34.8} && \textbf{32.2} & \textbf{41.1} & \textbf{34.4} & \textbf{36.1} \\
\Xhline{1.2pt}
VideoLLaMA (ZS) & 75.8 & 74.0 & 73.8 && 76.1 & 75.8 & 75.5 & 75.2 & \textbf{49.7} & \textbf{49.6} & \textbf{47.1} && \textbf{50.5} & \textbf{48.1} & \textbf{49.8} & \textbf{49.1} \\
\rowcolor{gray!10}VideoLLaMA (SFT)& \underline{76.2} & 73.4 & 73.6 && 76.0 & \underline{76.7} & 76.3 & 75.4 & \underline{47.1} & \underline{46.6} & \underline{45.6} && \underline{46.9} & 35.0 & \underline{45.9} & \underline{44.5} \\
VITA (ZS) & 73.2 & \underline{76.6} & \underline{76.8} && \underline{76.9} & \underline{76.7} & \underline{76.7} & \underline{76.1} & 29.6 & 31.3 & 31.9 && 31.4 & 31.8 & 31.8 & 31.3 \\
\rowcolor{gray!10}VITA (SFT) & \textbf{82.0} & \textbf{83.4} & \textbf{83.6} && \textbf{83.6} & \textbf{83.6} & \textbf{83.6} & \textbf{83.3} & 37.7 & {41.1} & {41.8} && {41.1} & \underline{44.4} & 42.2 & 41.4 \\
\Xhline{1.2pt}
\end{tabular}}
\vspace{-1mm}
\caption{Prediction accuracies (in \%) on Kinetics50-C (left) and VGGSound-C (right) datasets (with distribution shifts on the audio modality). We \textbf{bold} the best results and \underline{underline} the second-best.}

\label{tab:audio}
\end{table*}

\begin{table*}[ht]
\centering
\resizebox{\textwidth}{!}{
\begin{tabular}{l ccc c cccc c cccc c cccc c}
\Xhline{1.2pt}
\rowcolor{tableheadcolor!20} & \multicolumn{3}{c}{Noise} && \multicolumn{4}{c}{Blur} && \multicolumn{4}{c}{Weather} && \multicolumn{4}{c}{Digital} &  \\
\cline{2-4} \cline{6-9} \cline{11-14} \cline{16-19}\rowcolor{tableheadcolor!20}\multirow{-2}{*}{Models} & Gauss. & Shot & Impul. && Defoc. & Glass & Mot. & Zoom && Snow & Frost & Fog & Brit. && Contr. & Elas. & Pix. & JPEG &\multirow{-2}{*}{Avg.}\\
\Xhline{1.2pt}
\rowcolor{gray!10}CAV-MAE & 46.8 & 48.0 & 46.9 && 67.5 & 62.2 & 70.6 & 67.7 && 61.6 & 60.3 & 46.7 & 75.2 && 52.1 & 65.7 & 66.5 & 61.9 & 59.9 \\
\quad +MMT & 46.2 & 46.6 & 46.1 && 58.8 & 55.7 & 62.4 & 61.7 && 52.6 & 54.4 & 48.5 & 69.3 && 49.3 & 57.6 & 56.4 & 54.5 & 54.5 \\
\rowcolor{gray!10}\quad +Tent & 46.3 & 47.0 & 46.3 && 67.4 & 62.5 & 70.4 & 67.7 && 63.1 & 61.1 & 34.9 & 75.4 && 51.6 & 66.7 & 66.5 & 62.0 & 59.4 \\
\quad +EATA & 46.8 & 47.6 & 47.1 && 67.2 & 61.8 & 70.2 & 67.7 && 61.6 & 60.6 & 46.0 & 75.2 && 52.4 & 65.9 & 66.4 & 62.7 & 60.1 \\
\rowcolor{gray!10}\quad +SAR & 46.7 & 47.4 & 46.6 && 67.0 & 61.7 & 70.0 & 66.4 && 61.8 & 60.6 & 46.0 & 75.2 && 52.1 & 65.7 & 66.0 & 62.0 & 59.8 \\
\quad +READ & \underline{49.4} & \underline{49.7} & \underline{49.0} && \underline{68.0} & \underline{65.1} & \underline{71.2} & \underline{69.0} && \underline{64.5} & \underline{64.4} & \underline{57.4} & \underline{75.5} && \underline{53.6} & \underline{68.3} & \underline{68.0} & \underline{65.1} & \underline{62.5} \\
\rowcolor{gray!10}\quad +{ABPEM} & \textbf{50.3} & \textbf{51.1} & \textbf{50.4} && \textbf{70.0} & \textbf{69.6} & \textbf{72.5} & \textbf{71.2} && \textbf{65.2} & \textbf{66.2} & \textbf{65.6} & \textbf{75.7} && \textbf{56.6} & \textbf{71.9} & \textbf{70.5} & \textbf{67.8} & \textbf{65.0} \\
\Xhline{1.2pt}

VideoLLaMA (ZS) & \underline{23.8} & \underline{25.0} & \underline{25.8} && 39.6 & 32.7 & 39.3 & 42.9 && 40.8 & 35.2 & 47.9 & 60.7 && \underline{34.6} & 37.9 & \underline{57.7} & 49.4 & 39.5 \\
\rowcolor{gray!10}VideoLLaMA (SFT) & \textbf{26.6} & \textbf{27.9} & \textbf{29.6} && \textbf{46.4} & \underline{36.9} & \underline{45.1} & \underline{48.4} && \underline{45.6} & 38.8 & \underline{53.0} & \underline{67.0} && \textbf{39.4} & \textbf{42.1} & \textbf{64.9} & \underline{55.1} & \textbf{44.4} \\
VITA (ZS) & 14.3 & 14.7 & 16.1 && 30.7 & 33.0 & 39.7 & 43.5 && 36.5 & \underline{41.4} & 44.3 & 60.2 && 14.1 & 30.7 & 40.7 & 49.8 & 34.0 \\
\rowcolor{gray!10}VITA (SFT) & 20.5 & 21.1 & 23.0 && \underline{41.6} & \textbf{45.1} & \textbf{48.9} & \textbf{54.3} && \textbf{47.2} & \textbf{51.1} & \textbf{54.8} & \textbf{72.1} && 17.6 & \underline{41.5} & 54.0 & \textbf{59.9} & \underline{43.5} \\
\Xhline{1.2pt}
\end{tabular}}
\vspace{-1mm}
\caption{Prediction accuracies (in \%) on Kinetics50-C dataset (with distribution shifts on the visual modality). We \textbf{bold} the best results and \underline{underline} the second-best.}

\label{tab:ks50v}
\end{table*}

\subsection{Generalizability}
We then investigate how MLLMs generalize under test-time distribution shifts. Specifically, we adopt 15 types of distribution shifts on the visual modality (\emph{i.e.}, "Gaussian Noise", "Impulse Noise", "Shot Noise", "Glass Blur", "Defocus Blur", "Zoom Blur", "Motion Blur", "Snow", "Fog", "Frost", "Brightness", "Contrast", "Pixelate", "Elastic", and "JPEG Compression") and 6 types of distribution shifts on the audio modality (\emph{i.e.}, "Gaussian Noise", "Crowd Noise", "Traffic Noise", "Rain Noise", "Wind Noise" and "Thunder Noise") \cite{hendrycks2019benchmarking, yang2024test}. Examples of the distribution shifts are shown in Figure \ref{fig:corruption-vis}. We evaluate the models' performance under these distribution shifts at test time, comparing various test-time distribution methods (\emph{e.g.}, MMT, Tent, etc.) that are designed for traditional models to mitigate the distribution shifts, and the results are shown in Table \ref{tab:ks50v}, Table \ref{tab:audio}, and Table \ref{tab:vggv}.

\begin{table*}[h!]
\centering
\resizebox{\textwidth}{!}{
\begin{tabular}{l ccc c cccc c cccc c cccc c}
    \Xhline{1.2pt}
\rowcolor{tableheadcolor!20} & \multicolumn{3}{c}{Noise} && \multicolumn{4}{c}{Blur} && \multicolumn{4}{c}{Weather} && \multicolumn{4}{c}{Digital} &  \\
\cline{2-4} \cline{6-9} \cline{11-14} \cline{16-19}\rowcolor{tableheadcolor!20}\multirow{-2}{*}{Models} & Gauss. & Shot & Impul. && Defoc. & Glass & Mot. & Zoom && Snow & Frost & Fog & Brit. && Contr. & Elas. & Pix. & JPEG &\multirow{-2}{*}{Avg.}\\
\Xhline{1.2pt}
\rowcolor{gray!10}CAV-MAE       & 52.8  & 52.7  & 52.7  && 57.2  & 57.2  & 58.7  & 56.8  && 56.4  & 56.6  & 55.6  & 58.9  && 53.7  & 56.9  & 55.8  & 56.9  & 56.0 \\
\quad +MMT               & 7.1   & 7.3   & 7.3   && 44.8  & 41.5  & 48.0  & 45.5  && 27.4  & 23.5  & 30.5  & 46.3  && 24.0  & 43.0  & 40.7  & 45.7  & 32.0 \\
\rowcolor{gray!10}\quad +Tent   & 52.7  & 52.7  & 52.7  && 56.7  & 56.5  & 58.0  & 56.5  && 55.0  & 57.0  & 56.3  & 58.7  && 54.0  & 57.4  & 56.7  & 57.4  & 55.8 \\
\quad +EATA              & 53.0  & 52.8  & 53.0  && 57.2  & 57.1  & 58.6  & 57.8  && 56.3  & 56.8  & \underline{56.4}  & 59.0  && 54.1  & 57.4  & 56.1  & 57.0  & 56.2 \\
\rowcolor{gray!10}\quad +SAR    & 52.9  & 52.8  & 52.9  && 57.0  & 57.1  & 58.5  & 56.8  && 56.3  & 56.7  & 55.9  & 58.9  && 54.0  & 57.6  & \underline{57.1}  & 57.2  & 56.1 \\
\quad +READ              & \underline{53.6}  & \underline{53.6}  & \underline{53.5}  && \underline{57.9}  & \underline{57.7}  & \underline{59.4}  & \underline{58.8}  && \underline{57.2}  & \underline{57.8}  & 55.0  & \underline{59.9}  && \underline{55.2}  & \underline{58.6}  & \underline{57.1}  & \underline{57.9}  & \underline{56.9} \\
\rowcolor{gray!10}\quad +{ABPEM}  & \textbf{54.0}  & \textbf{53.9}  & \textbf{54.0}  && \textbf{58.2}  & \textbf{58.1}  & \textbf{59.6}  & \textbf{59.3}  && \textbf{57.5}  & \textbf{58.2}  & \textbf{58.2}  & \textbf{60.2}  && \textbf{56.2}  & \textbf{59.1}  & \textbf{57.5}  & \textbf{58.3}  & \textbf{57.5} \\
\Xhline{1.2pt}

VideoLLaMA (ZS)         & \underline{39.1}  & \underline{39.5}  & \underline{39.6}  && \underline{48.0}  & \underline{44.1}  & \underline{47.4}  & \underline{47.4}  && \underline{36.6}  & \underline{39.9}  & \underline{48.4}  & \underline{54.0}  && \underline{45.8}  & \underline{43.3}  & \underline{53.3}  & \underline{50.8}  & \underline{45.1} \\
\rowcolor{gray!10}VideoLLaMA (SFT)& \textbf{46.8}  & \textbf{47.2}  & \textbf{47.5}  && \textbf{52.8}  & \textbf{49.6}  & \textbf{52.9}  & \textbf{53.6}  && \textbf{46.7}  & \textbf{49.3}  & \textbf{54.6}  & \textbf{59.7}  && \textbf{52.6}  & \textbf{50.3}  & \textbf{56.9}  & \textbf{56.3}  & \textbf{51.8} \\
VITA (ZS)                & 5.9   & 6.4   & 6.4   && 11.6  & 11.4  & 13.9  & 14.3  && 13.9  & 17.5  & 19.2  & 23.3  && 5.3   & 10.9  & 14.5  & 17.4  & 12.8 \\
\rowcolor{gray!10}VITA (SFT)& 13.1  & 13.0  & 14.4  && 16.7  & 16.5  & 18.7  & 17.4  && 16.1  & 18.1  & 20.6  & 25.6  && 12.9  & 14.3  & 21.6  & 21.6  & 17.4 \\
\Xhline{1.2pt}
\end{tabular}}
\vspace{-1mm}
\caption{Prediction accuracies (in \%) on VGGSound-C dataset (with distribution shifts on the visual modality). We \textbf{bold} the best results and \underline{underline} the second-best.}
\vspace{-1mm}
\label{tab:vggv}
\end{table*}

    \begin{table*}[ht]
    \centering
    \resizebox{0.95\textwidth}{!}{
    \begin{tabular}{l ccccc c ccccc}
    \Xhline{1.2pt}
    \rowcolor{tableheadcolor!20} & \multicolumn{5}{c}{Kinetics50} && \multicolumn{5}{c}{VGGSound} \\
    \cline{2-6} \cline{8-12}\rowcolor{tableheadcolor!20}\multirow{-2}{*}{Models} & Clean & FGSM & ASR & PGD & ASR && Clean & FGSM & ASR & PGD & ASR \\
    \Xhline{1.2pt}
    \rowcolor{gray!10}CAV-MAE & \underline{82.3} & 43.2 & \textbf{47.5\%} & 31.4 & \textbf{61.8\%} && \textbf{65.5} & 39.1 & \textbf{40.2\%} & 36.3 & \textbf{44.6\%}  \\
    Video-LLaMA2 (Zero-Shot) & 73.2 & 72.8 & 0.6\% & 72.5 & 0.9\% && 59.3 & \underline{59.2} & 0.2\% & \underline{58.6} & 1.2\% \\
    \rowcolor{gray!10}Video-LLaMA2 (SFT) & 78.9 & \underline{77.4} & \underline{1.8\%}  & \underline{77.3} & \underline{1.9\%} && \underline{63.1} & \textbf{61.0} & \underline{3.3\%} & \textbf{61.8} & \underline{2.1\%} \\
    VITA (Zero-Shot) & 70.5 & 70.1 & 0.6\% & 70.2 & 0.4\% && 29.8 & 29.3 & 1.8\% & 29.2 & 2.0\% \\
    \rowcolor{gray!10}VITA (SFT) & \textbf{84.3} & \textbf{83.6} & 0.9\% & \textbf{84.1} & 0.3\% && 32.0 & 31.6 & 1.4\% & 31.3 & 2.1\% \\
    \Xhline{1.2pt}
    \end{tabular}}
    \vspace{-1mm}
    \caption{Models' performance under adversarial attacks. We \textbf{bold} the best results and \underline{underline} the second-best.}
    \vspace{-2mm}
    \label{tab:adv}
\end{table*}

\obs{Observation 6: MLLMs are prone to test-time distribution shifts in the visual modality.}
As can be seen from the results, test-time distribution shifts on the visual modality generally lead to a significant performance degradation for MLLMs, while the performance degradation on the audio modality is less severe. This can be attributed to the MLLMs' over-reliance on the visual modality (as discussed in Section \ref{sec:effective}), which makes them vulnerable to distribution shifts on the input video. We also find that when the audio modality is corrupted at test time, there is an increase in the VITA's performance, which is consistent with the observation of the negative synergistic effect between the modalities (as discussed in Section \ref{sec:effective}). Moreover, we find that previous test-time adaptation solutions are problem-specific (specially designed for the classification problem with entropy-based objectives) and architecture-specific (specially designed for models with certain architectures). The performance degradation of MLLMs, especially under visual distribution shifts, calls for new solutions to improve their generalizability.

\subsection{Robustness Against Adversarial Perturbations}
In this part, we evaluate the robustness of MLLMs' audio-visual capabilities against adversarial perturbations. We adopt two commonly used adversarial attack methods, \emph{i.e.}, FGSM \cite{goodfellow2014explaining} and PGD \cite{madry2017towards}, to generate adversarial examples for the models. Specifically, as the audio signals are processed with non-differentiable operations, we only attack the visual modality. For fast gradient sign method (FGSM), we use the following equation:
\begin{equation}
\label{eq:fgsm}
    \tilde{\bm X}^V = \bm X^V + \epsilon \cdot \text{sign}(\nabla_{\bm X^V} \mathcal{L}_{\text{CE}}),
\end{equation}
where $\epsilon$ is the perturbation magnitude, and $\mathcal{L}_{\text{CE}}$ is the cross-entropy loss function. We set $\epsilon$ to 0.01. For projected gradient descent (PGD), we use the following equation:
\begin{equation}
    \label{eq:pgd}
    \tilde{\bm X}^V = \Pi_\epsilon(\bm X^V + \alpha \cdot \nabla_{\bm X^V} \mathcal{L}_{\text{CE}}),
\end{equation}
where $\alpha$ is the step size. Eq. \ref{eq:pgd} is computed iteratively (we perform 10 iterations in this paper). We set $\alpha$ to 0.5, and $\epsilon$ to 0.01.
We evaluate the models' performance under these adversarial examples, and the results are shown in Table \ref{tab:adv}, where we also report the attack success rate (ASR, \citet{eykholt2018robust}). \obs{Observation 8: MLLMs are robust against adversarial attacks.} As can be seen from the results, MLLMs are generally robust against adversarial perturbations compared to traditional models, with the attack success rate being much lower than that of CAV-MAE. This may be attributed to the MLLMs' audio-visual capability and its integration with LLMs. The complexity of the language model makes it difficult for attackers to perform black-box attacks against MLLMs. Thus, for closed-source MLLMs, performing effective adversarial attacks is challenging.

\begin{figure}[t]
    \centering
    \includegraphics[width=\linewidth]{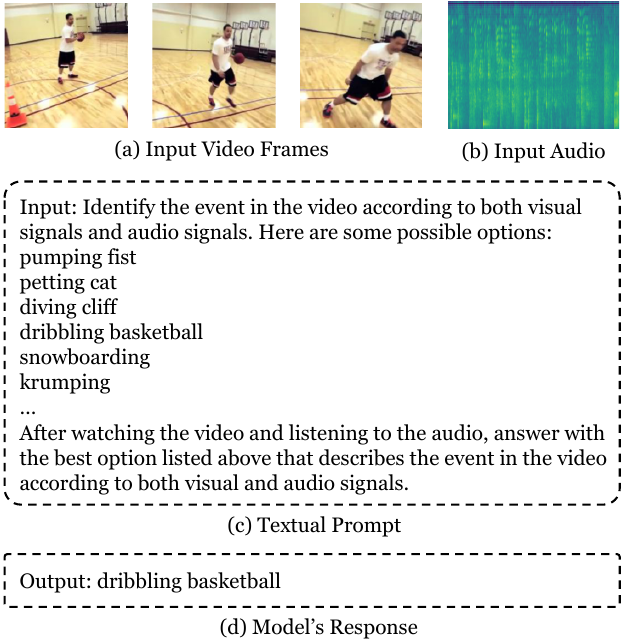}
    \vspace{-5mm}
    \caption{An example where the model generates the correct answer. The input video frames and audio signals are shown in subfigures (a) and (b), the textual prompt is shown in subfigure (c) and the model's output is shown in subfigure (d).}
    \vspace{-5mm}
    \label{fig:case1}
\end{figure}

\begin{figure}[t]
    \centering
    \includegraphics[width=\linewidth]{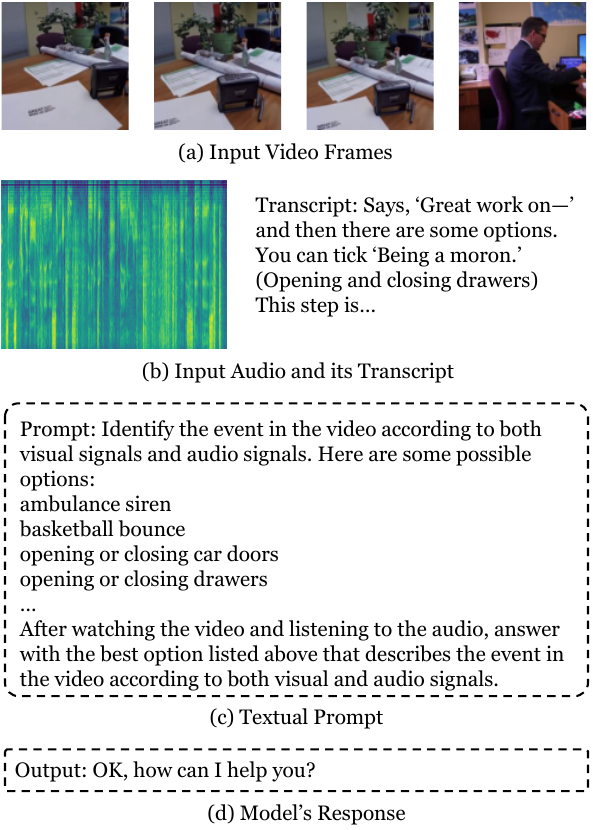}
    \vspace{-6mm}
    \caption{An example of the model's confusion between speech and textual instructions. We also show the transcript of the audio signals in subfigure (b).}
    \vspace{-5mm}
    \label{fig:case2}
\end{figure}

\begin{figure}[t]
    \centering
    \includegraphics[width=\linewidth]{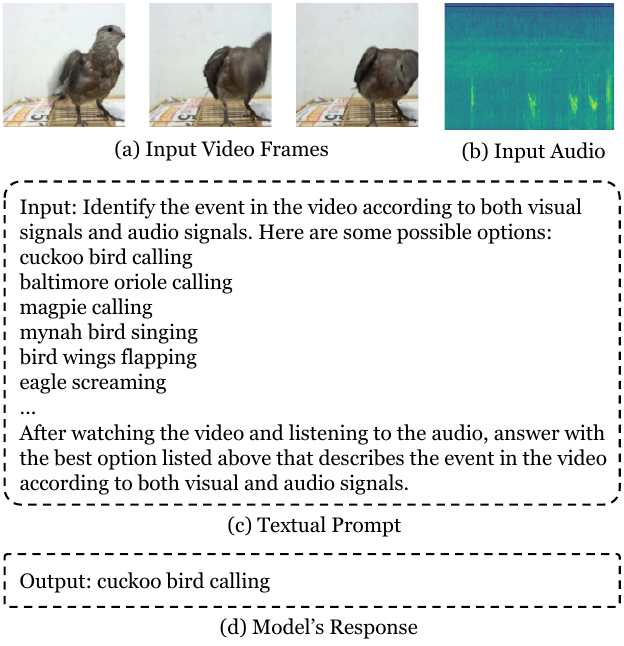}
    \vspace{-6mm}
    \caption{An example of the model's over-reliance on the visual modality. We show that visually similar birds may have different sounds.}
    \vspace{-5mm}
    \label{fig:case3}
\end{figure}

\subsection{Case Study}
\label{sec:case}
In this section, we provide specific cases of the models' outputs given specific inputs. 

\noindent\textit{\textbf{Case 1: Correct Answer Prediction.}}
We first show an example where the model correctly predicts the answer in Figure \ref{fig:case1}. In this case, the correct output can be directly inferred from the input video frames, where we can see a man dribbling a basketball (Figure \ref{fig:case1}a). The audio signal is also informative, as we hear the sound of a basketball bouncing (Figure \ref{fig:case1}b, although it is not clear from the visualization of audio signals). The model's output is consistent with the input, demonstrating the model's ability to understand the audio-visual information.

\noindent\textit{\textbf{Case 2: Confusion Between Speech and Textual Instructions.}}
We then show an example where the model is confused between speech and textual instructions in Figure \ref{fig:case2}. In this case, the input video frames (Figure \ref{fig:case2}a) show a man sitting at an office desk with papers and a computer screen. The input audio contains both speech and other sounds (Figure \ref{fig:case2}b). The man seems to be filling out a table while speaking, during which he opens and closes the drawer. The textual prompt (Figure \ref{fig:case2}c) asks the MLLM to identify the event based on the video and audio. However, the model seems to ignore these textual instructions, and instead asks what it can do for the man in the video (Figure \ref{fig:case2}d). This suggests that the model is confused with the speech and textual instructions. The audio signals, while from a different modality, carry the information that plays a similar role as the text (\emph{i.e.}, providing instructions), and the model takes the instructions from the audio, ignoring initial textual instructions.

\noindent\textit{\textbf{Case 3: Over-Reliance on the Visual Modality.}}
We have previously mentioned that current MLLMs tend to over-rely on the visual modality while ignoring the audio modality, which can be problematic when the audio modality carries important information. We provide an example of this over-reliance in Figure \ref{fig:case3}. In this case, the input video frames (Figure \ref{fig:case3}a) show a little bird jumping around on a cardboard box. The audio signal (Figure \ref{fig:case3}b) contains the sound of this bird. The textual instruction (Figure \ref{fig:case3}c) requires the MLLM to differentiate the type of this bird. Some birds are visually similar (\emph{e.g.} cuckoo bird and mynah bird), but their sounds are different. The model's output (Figure \ref{fig:case3}d) is incorrect, as it fails to match the sound of the bird in the input audio (in this case, the sound of a mynah bird) with the visual information. This demonstrates the model's over-reliance on the dominant modality (vision) can lead to problems when the other modality (audio) is critical.

\section{Conclusion}
\label{sec:conclusion}
This paper evaluates the audio-visual capabilities of MLLMs across four key dimensions: effectiveness, efficiency, generalizability, and robustness. The results show that MLLMs are generally effective in understanding audio-visual information, although they rely heavily on the visual modality, which leads to poor performance when video inputs undergo test-time distribution shifts. In addition, MLLMs exhibit high data efficiency with superior performance under limited data, but they lag behind in terms of computational efficiency. Furthermore, MLLMs are more robust compared to traditional models against adversarial attacks. These findings highlight the strengths and limitations of current MLLMs in handling audio-visual information, providing comprehensive evaluations and offering guidance for future research.

\section*{Limitations}
Despite extensive evaluations, we should note that this paper does not involve solutions to the problems presented, including over-reliance on the visual modality, weak generalizability when the visual modality is under distribution shifts, and the high computational cost of MLLMs. Future work should focus on addressing these limitations to improve the audio-visual capabilities of MLLMs.

\bibliography{acl_latex}

\end{document}